\pdfoutput=1
\documentclass[12pt,a4paper]{article}
\usepackage{ifthen} % for conditional statements
\newboolean{pdflatex}
\setboolean{pdflatex}{true} % False for eps figures 

\newboolean{articletitles}
\setboolean{articletitles}{true} % False removes titles in references

\newboolean{uprightparticles}
\setboolean{uprightparticles}{false} %True for upright particle symbols

\newboolean{inbibliography}
\setboolean{inbibliography}{false} %True once you enter the bibliography

\def\paperauthors{LHCb collaboration} % Leave as is for PAPER, CONF and FIGURE
\def\paperasciititle{Observation of the rare baryonic decay B+ -> p antiLambda and measurement of its weak decay parameter} % Set ASCII title here !! MAKE sure it's only ASCII characters !! 
\def\papertitle{Observation of the Rare Baryonic Decay \BuPLbar and Measurement of its Weak Decay Parameter} % Latex formatted title
\def\paperkeywords{{High Energy Physics}, {LHCb}} % Comma separated list
\def\papercopyright{\the\year\ CERN for the benefit of the LHCb collaboration} % new since 9/Apr/2018
\def\paperlicence{CC BY 4.0 licence}
\def\paperlicenceurl{https://creativecommons.org/licenses/by/4.0/}

\newif\ifEnableSectionTOCLinks
\EnableSectionTOCLinksfalse % deactivated

\usepackage[top=1in, bottom=1.25in, left=1in, right=1in]{geometry}

\usepackage{microtype}
\usepackage{lineno}  % for line numbering during review
\usepackage{xspace} % To avoid problems with missing or double spaces after
\usepackage{caption} %these three command get the figure and table captions automatically small

\usepackage{graphicx}  % to include figures (can also use other packages)
\usepackage{color}
\usepackage{colortbl}
\graphicspath{{./figs/}} % Make Latex search fig subdir for figures
\usepackage{amsmath} % Adds a large collection of math symbols
\usepackage{amssymb}
\usepackage{amsfonts}
\usepackage{upgreek} % Adds in support for greek letters in roman typeset

\newcommand*\patchAmsMathEnvironmentForLineno[1]{%
\expandafter\let\csname old#1\expandafter\endcsname\csname #1\endcsname
\expandafter\let\csname oldend#1\expandafter\endcsname\csname
end#1\endcsname
 \renewenvironment{#1}%
   {\linenomath\csname old#1\endcsname}%
   {\csname oldend#1\endcsname\endlinenomath}%
}
\newcommand*\patchBothAmsMathEnvironmentsForLineno[1]{%
  \patchAmsMathEnvironmentForLineno{#1}%
  \patchAmsMathEnvironmentForLineno{#1*}%
}
\AtBeginDocument{%
\patchBothAmsMathEnvironmentsForLineno{equation}%
\patchBothAmsMathEnvironmentsForLineno{align}%
\patchBothAmsMathEnvironmentsForLineno{flalign}%
\patchBothAmsMathEnvironmentsForLineno{alignat}%
\patchBothAmsMathEnvironmentsForLineno{gather}%
\patchBothAmsMathEnvironmentsForLineno{multline}%
\patchBothAmsMathEnvironmentsForLineno{eqnarray}%
}

\usepackage[pdftex,
            pdfauthor={\paperauthors},
            pdftitle={\paperasciititle},
            pdfkeywords={\paperkeywords}]{hyperref}
\usepackage{hyperxmp}
\hypersetup{
    pdfcopyright={Copyright (C) \papercopyright},
    pdflicenseurl={\paperlicenceurl}
}
\usepackage[colorinlistoftodos,textsize=scriptsize]{todonotes}

\usepackage[bottom,flushmargin,hang,multiple]{footmisc}

\usepackage[all]{hypcap} % Internal hyperlinks to floats.

\usepackage{xspace} 
\usepackage{upgreek}

\def\lhcb   {\mbox{LHCb}\xspace}

\def\babar  {\mbox{BaBar}\xspace}
\def\belle  {\mbox{Belle}\xspace}

\def\cleo   {\mbox{CLEO}\xspace}

\def\MagUp {\mbox{\em Mag\kern -0.05em Up}\xspace}

\ifthenelse{\boolean{uprightparticles}}%
{

 \def\Ppi         {\ensuremath{\uppi}\xspace}

 \def\Ppsi        {\ensuremath{\uppsi}\xspace}

 \def\PDelta      {\ensuremath{\Delta}\xspace}                 
 \def\PXi         {\ensuremath{\Xi}\xspace}                 
 \def\PLambda     {\ensuremath{\Lambda}\xspace}                 
 \def\PSigma      {\ensuremath{\Sigma}\xspace}                 
 \def\POmega      {\ensuremath{\Omega}\xspace}                 
 \def\PUpsilon    {\ensuremath{\Upsilon}\xspace}
 
 \def\PPi         {\ensuremath{\oldPi}\xspace}

 \def\PB      {\ensuremath{\mathrm{B}}\xspace}                 
 \def\PD      {\ensuremath{\mathrm{D}}\xspace}                 

 \def\PK      {\ensuremath{\mathrm{K}}\xspace}                 
 \def\Pb      {\ensuremath{\mathrm{b}}\xspace}                 
                  
 \def\Pd      {\ensuremath{\mathrm{d}}\xspace}                 

 \def\Pp      {\ensuremath{\mathrm{p}}\xspace}                 

 \def\Ps      {\ensuremath{\mathrm{s}}\xspace}                 
                  
 \def\Pu      {\ensuremath{\mathrm{u}}\xspace}                 
 \def\thebaroffset{0.0em}
}
{

 \def\Ppi         {\ensuremath{\pi}\xspace}

 \def\Ppsi        {\ensuremath{\psi}\xspace}                 
                  
 \mathchardef\PDelta="7101
 \mathchardef\PXi="7104
 \mathchardef\PLambda="7103
 \mathchardef\PSigma="7106
 \mathchardef\POmega="710A
 \mathchardef\PUpsilon="7107
 \mathchardef\PPi="7105
 \def\PB      {\ensuremath{B}\xspace}                 
 \def\PD      {\ensuremath{D}\xspace}                 

 \def\PK      {\ensuremath{K}\xspace}                 
 \def\Pb      {\ensuremath{b}\xspace}                 
                  
 \def\Pd      {\ensuremath{d}\xspace}                 

 \def\Pp      {\ensuremath{p}\xspace}                 

 \def\Ps      {\ensuremath{s}\xspace}                 
                  
 \def\Pu      {\ensuremath{u}\xspace}                 
 \def\thebaroffset{0.18em}
}
\newcommand{\offsetoverline}[2][\thebaroffset]{\kern #1\overline{\kern -#1 #2}}%

\makeatletter
\ifcase \@ptsize \relax% 10pt
  \newcommand{\miniscule}{\@setfontsize\miniscule{4}{5}}% \tiny: 5/6
\or% 11pt
  \newcommand{\miniscule}{\@setfontsize\miniscule{5}{6}}% \tiny: 6/7
\or% 12pt
  \newcommand{\miniscule}{\@setfontsize\miniscule{5}{6}}% \tiny: 6/7
\fi
\makeatother

\DeclareRobustCommand{\optbar}[1]{\shortstack{{\miniscule (\rule[.5ex]{1.25em}{.18mm})}
  \\ [-.7ex] $#1$}}

\def\uquark    {{\ensuremath{\Pu}}\xspace}

\def\dquark    {{\ensuremath{\Pd}}\xspace}

\def\squark    {{\ensuremath{\Ps}}\xspace}

\def\bquark    {{\ensuremath{\Pb}}\xspace}

\def\pion   {{\ensuremath{\Ppi}}\xspace}
\def\piz    {{\ensuremath{\pion^0}}\xspace}
\def\pip    {{\ensuremath{\pion^+}}\xspace}
\def\pim    {{\ensuremath{\pion^-}}\xspace}

\def\kaon    {{\ensuremath{\PK}}\xspace}
\def\KorKbar {\kern \thebaroffset\optbar{\kern -\thebaroffset \PK}{}\xspace}

\def\Kp      {{\ensuremath{\kaon^+}}\xspace}
\def\Km      {{\ensuremath{\kaon^-}}\xspace}

\def\KS      {{\ensuremath{\kaon^0_{\mathrm{S}}}}\xspace}

\def\Kstar   {{\ensuremath{\kaon^*}}\xspace}

\def\D       {{\ensuremath{\PD}}\xspace}

\def\DorDbar {\kern \thebaroffset\optbar{\kern -\thebaroffset \PD}\xspace}
\def\Dz      {{\ensuremath{\D^0}}\xspace}

\def\Dp      {{\ensuremath{\D^+}}\xspace}
\def\Dm      {{\ensuremath{\D^-}}\xspace}

\def\DpDm    {\ensuremath{\Dp {\kern -0.16em \Dm}}\xspace}

\def\B       {{\ensuremath{\PB}}\xspace}

\def\BorBbar {\kern \thebaroffset\optbar{\kern -\thebaroffset \PB}\xspace}

\def\Bd      {{\ensuremath{\B^0}}\xspace}

\def\BdorBdbar {\kern \thebaroffset\optbar{\kern -\thebaroffset \Bd}\xspace}
\def\Bu      {{\ensuremath{\B^+}}\xspace}

\def\Bp      {{\ensuremath{\Bu}}\xspace}

\def\Bs      {{\ensuremath{\B^0_\squark}}\xspace}

\def\BsorBsbar {\kern \thebaroffset\optbar{\kern -\thebaroffset \Bs}\xspace}

\def\Bds     {{\ensuremath{\B_{(\squark)}^0}}\xspace}

\def\psitwos  {{\ensuremath{\Ppsi{(2S)}}}\xspace}

\def\Y#1S{\ensuremath{\PUpsilon{(#1S)}}\xspace}

\def\proton      {{\ensuremath{\Pp}}\xspace}
\def\antiproton  {{\ensuremath{\overline \proton}}\xspace}

\def\Lbar        {{\ensuremath{\offsetoverline{\PLambda}}}\xspace}
\def\LorLbar     {\kern \thebaroffset\optbar{\kern -\thebaroffset \PLambda}\xspace}

\def\BF         {{\ensuremath{\mathcal{B}}}\xspace}
\def\BR         {\BF}

\newcommand{\decay}[2]{\mbox{\ensuremath{#1\!\to #2}}\xspace} 

\def\to                 {\ensuremath{\rightarrow}\xspace}

\def\CP                {{\ensuremath{C\!P}}\xspace}

\def\AT#1     {\ensuremath{A_{\mathrm{T}}^{#1}}\xspace}           % 2

\def\C#1      {\ensuremath{\mathcal{C}_{#1}}\xspace}                       % 9
\def\Cp#1     {\ensuremath{\mathcal{C}_{#1}^{'}}\xspace}                    % 7
\def\Ceff#1   {\ensuremath{\mathcal{C}_{#1}^{\mathrm{(eff)}}}\xspace}        % 9  
\def\Cpeff#1  {\ensuremath{\mathcal{C}_{#1}^{'\mathrm{(eff)}}}\xspace}       % 7
\def\Ope#1    {\ensuremath{\mathcal{O}_{#1}}\xspace}                       % 2
\def\Opep#1   {\ensuremath{\mathcal{O}_{#1}^{'}}\xspace}                    % 7

\newcommand{\aunit}[1]{\ensuremath{\text{\,#1}}}       
\newcommand{\tev}{\aunit{Te\kern -0.1em V}\xspace}
\newcommand{\gev}{\aunit{Ge\kern -0.1em V}\xspace}
\newcommand{\mev}{\aunit{Me\kern -0.1em V}\xspace}
\newcommand{\kev}{\aunit{ke\kern -0.1em V}\xspace}
\newcommand{\ev}{\aunit{e\kern -0.1em V}\xspace}
 
\newcommand{\mevc}{\ensuremath{\aunit{Me\kern -0.1em V\!/}c}\xspace}
\newcommand{\gevc}{\ensuremath{\aunit{Ge\kern -0.1em V\!/}c}\xspace}
\newcommand{\mevcc}{\ensuremath{\aunit{Me\kern -0.1em V\!/}c^2}\xspace}
\newcommand{\gevcc}{\ensuremath{\aunit{Ge\kern -0.1em V\!/}c^2}\xspace}
\def\fb   {\ensuremath{\aunit{fb}}\xspace}
\def\invfb   {\ensuremath{\fb^{-1}}\xspace}

\newcommand{\chisq}{\ensuremath{\chi^2}\xspace}

\newcommand{\chisqip}{\ensuremath{\chi^2_{\text{IP}}}\xspace}

\def\deriv {\ensuremath{\mathrm{d}}}

\def\gsim{{~\raise.15em\hbox{$>$}\kern-.85em
          \lower.35em\hbox{$\sim$}~}\xspace}
\def\lsim{{~\raise.15em\hbox{$<$}\kern-.85em
          \lower.35em\hbox{$\sim$}~}\xspace}

\newcommand{\Real}{\ensuremath{\mathcal{R}e}\xspace}

\def\tell1  {TELL1\xspace}
\def\ukl1   {UKL1\xspace}

\newcommand{\eg}{\mbox{\itshape e.g.}\xspace}

\newcommand{\lhcborcid}[1]{\href{https://orcid.org/#1}{\hspace*{0.1em}\raisebox{-0.45ex}{\includegraphics[width=1em]{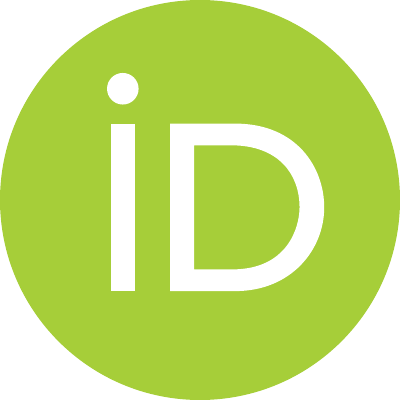}}}}

\hypersetup{
  colorlinks   = true, %Colours links instead of ugly boxes
  urlcolor     = blue, %Colour for external hyperlinks
  linkcolor    = blue, %Colour of internal links
  citecolor    = red   %Colour of citations
}

\ifEnableSectionTOCLinks
    \usepackage[explicit]{titlesec} % to change headings
    
    \let\oldcontentsline\contentsline
    \renewcommand

    \titleformat{\section}{\normalfont\Large\bf}{\hyperlink{tocsection.\thesection}{{\thesection} \parbox[t]{\dimexpr\textwidth-1pc}{#1}}}{1pc}{}

    \titleformat{\subsection}{\normalfont\bf}{\hyperlink{tocsubsection.\thesubsection}{{\thesubsection} \parbox[t]{\dimexpr\textwidth-1pc}{#1}}}{1pc}{}

    \titleformat{name=\section,numberless}[display]{}{}{0pt}{\normalfont\Huge\bfseries #1}
\fi

\usepackage{cite} % Allows for ranges in citations
\usepackage{mciteplus}
\def\Bds{{\ensuremath{\B^0_{\kern -0.1em{\scriptscriptstyle (}\kern -0.05em\squark\kern -0.03em{\scriptscriptstyle )}}}}\xspace}

\def\Vz {\mbox{$V^0$}\xspace}

\def\Sbar{\kern 0.2em\overline{\kern -0.2em \Sigma}{}\xspace}

\def\psitwos  {\ensuremath{\Ppsi{(2S)}}\xspace}

\newcommand{\BuPLbar}{\texorpdfstring{\decay{\Bu}{\proton \Lbar}}{}}

\newcommand{\BuKSPi}{\texorpdfstring{\decay{\Bu}{\KS \pip}}{}}
\newcommand{\BuKSK}{\texorpdfstring{\decay{\Bu}{\KS \Kp}}{}}

\newcommand{\BdPPbar}{\texorpdfstring{\decay{\Bd}{\proton \antiproton}}{}}

\newcommand{\BdKPi}{\texorpdfstring{\decay{\Bd}{\Kp \pim}}{}}
\newcommand{\BdPiPi}{\texorpdfstring{\decay{\Bd}{\pip \pim}}{}}

\newcommand{\BsKK}{\texorpdfstring{\decay{\Bs}{\Kp \Km}}{}}

\newcommand{\BsPiK}{\texorpdfstring{\decay{\Bs}{\pip \Km}}{}}
\newcommand{\LbPPi}{\texorpdfstring{\decay{\PLambda_b^0}{\proton \pim}}{}}
\newcommand{\LbPK}{\texorpdfstring{\decay{\PLambda_b^0}{\proton \Km}}{}}
\newcommand{\BuPLbarExcited}{\texorpdfstring{\decay{\Bu}{\proton \Lbar(1520)}}{}}

\newcommand{\BdPLbarPi}{\texorpdfstring{\decay{\Bd}{\proton \Lbar \pim}}{}}

\newcommand{\BuPLbarPiz}{\texorpdfstring{\decay{\Bu}{\proton \Lbar \piz}}{}}

\newcommand{\BuPLbarGamma}{\texorpdfstring{\decay{\Bu}{\proton \Lbar \gamma}}{}}

\newcommand{\KSPiPi}{\texorpdfstring{\decay{\KS}{\pip \pim}}{}}
\newcommand{\LPPi}{\texorpdfstring{\decay{\PLambda}{\proton \pim}}{}}
\newcommand{\LbarPPi}{\texorpdfstring{\decay{\Lbar}{\antiproton \pip}}{}}
\newcommand{\Psipp}{\texorpdfstring{\psitwos \to \proton \antiproton}{}}

\def\Pbarpi {$\antiproton\pip$\xspace}

\def\PiPi  {$\pip\pim$\xspace}
\def\PPi   {$\proton\pim$\xspace}
\def\PLbar {\ensuremath{\proton\Lbar}\xspace}

\def\KSPi  {\ensuremath{\KS\pip}\xspace}

\def\KStarKpi   {\ensuremath{\Kstar(892)^0 \to \Kp \pim}\xspace}
\def\DKpi  {\ensuremath{\Dz\to\Km\pip}\xspace}

\def\splot{\mbox{\em sPlot}\xspace}

\def\thetap  {\ensuremath{\theta_{p}}\xspace}
\def\costhetap  {\ensuremath{\cos\theta_{p}}\xspace}
\def\alphaL  {\ensuremath{\alpha_{\Lambda}}\xspace}
\def\alphaB  {\ensuremath{\alpha_{B}}\xspace}

\newcommand{\btouus}{\texorpdfstring{\decay{\bar{\bquark}}{\uquark \bar{\uquark} \bar{\squark}}}{}}
\newcommand{\btouud}{\texorpdfstring{\decay{\bar{\bquark}}{\uquark \bar{\uquark} \bar{\dquark}}}{}}

\usepackage{multirow}
\usepackage{booktabs}
\usepackage{caption,subcaption}
\usepackage{setspace}
\usepackage{siunitx}

\begin{document}

%%%%%%%%%%%%%%%%%%%%%%%%%
%%%%% Title     %%%%%%%%%
%%%%%%%%%%%%%%%%%%%%%%%%%
\renewcommand{\thefootnote}{\fnsymbol{footnote}}
\setcounter{footnote}{1}

% %%%%%%% CHOOSE TITLE PAGE--------
%\onecolumn
%\input{title-LHCb-INT}
%\input{title-LHCb-ANA}
%\input{title-LHCb-CONF}
%\input{title-LHCb-FIGURE}
% ===============================================================================
% Purpose: LHCb-PAPER journal paper title page template
% Author: 
% Created on: 2010-09-25
% ===============================================================================

%%%%%%%%%%%%%%%%%%%%%%%%%
%%%%%  TITLE PAGE  %%%%%%
%%%%%%%%%%%%%%%%%%%%%%%%%
\begin{titlepage}
\pagenumbering{roman}

% Header ---------------------------------------------------
\vspace*{-1.5cm}
\centerline{\large EUROPEAN ORGANIZATION FOR NUCLEAR RESEARCH (CERN)}
\vspace*{1.5cm}
\noindent
\begin{tabular*}{\linewidth}{lc@{\extracolsep{\fill}}r@{\extracolsep{0pt}}}
%\ifthenelse{\boolean{pdflatex}}% Logo format choice
\vspace*{-1.5cm}\mbox{\!\!\!\includegraphics[width=.14\textwidth]{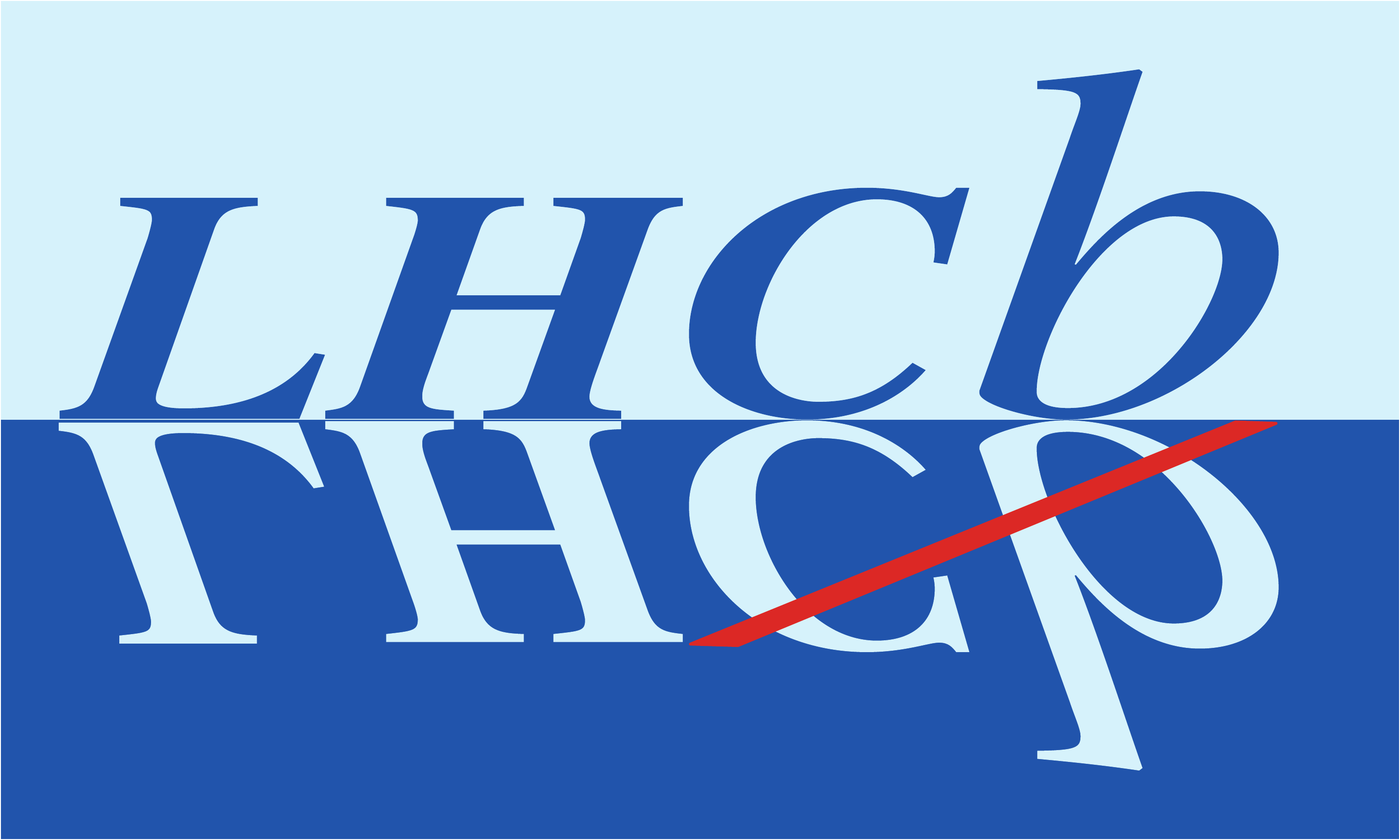}} & &
%{\vspace*{-1.2cm}\mbox{\!\!\!\includegraphics[width=.12\textwidth]{figs/lhcb-logo.eps}} & &}%
\\
 & & CERN-EP-2025-263 \\  % ID 
 & & LHCb-PAPER-2025-044 \\  % ID 
 & & February 7, 2026 \\ % Date - Can also hardwire e.g.: 23 March 2010
 & & \\
% not in paper \hline
\end{tabular*}

\vspace*{4.0cm}

% Title --------------------------------------------------
{\normalfont\bfseries\boldmath\huge
\begin{center}
% DO NOT EDIT HERE. Instead edit macro in main.tex to keep metadata correct
  \papertitle 
\end{center}
}

\vspace*{1.0cm}

% Authors -------------------------------------------------
\begin{center}
%In the footnote, replace 'paper' by 'Letter' in case of submission to PRL or PLB 
% Edit macro in main.tex to keep metadata correct
\paperauthors\footnote{Authors are listed at the end of this Letter.}
\end{center}
\vspace{\fill}
% Abstract -----------------------------------------------
\begin{abstract}
  \noindent
  The first observation of the decay $\Bp\rightarrow \proton \Lbar$ is presented, using proton-proton collision data collected by the LHCb experiment between 2016 and 2018 at a center-of-mass energy of $13\tev$, corresponding to an integrated luminosity of $5.4\invfb$. 
   The signal significance exceeds seven standard deviations.
   Using the $\Bp \rightarrow \KS \pip$  decay as a normalization channel, the branching fraction is measured and combined with previous LHCb results based on data collected at 7 and $8\tev$ 
  in 2011 and 2012,  yielding 
  \begin{eqnarray}
  \BF(\Bp\rightarrow \proton \Lbar) = (1.24\pm0.17\pm0.05\pm0.03)\times10^{-7}\,,\nonumber
  \end{eqnarray}
where the first uncertainty is statistical, the second is systematic, and the third comes from the uncertainty on the branching fraction of the normalization channel.
The $\Bp\rightarrow \proton \Lbar$ weak decay parameter is measured to be $\alpha_B = 0.87_{-0.29}^{+0.26} \pm 0.09$, indicating the presence of comparable S-wave and P-wave decay amplitudes. 
\end{abstract}

\vspace*{2.0cm}

\begin{center}
  Published in Physical Review Letters 136 051802 (2026).
\end{center}

\vspace{\fill}

{\footnotesize 
% Edit macro in main.tex to keep metadata correct
\centerline{\copyright~\papercopyright. \href{\paperlicenceurl}{\paperlicence}.}}
\vspace*{2mm}

\end{titlepage}

%%%%%%%%%%%%%%%%%%%%%%%%%%%%%%%%
%%%%%  EOD OF TITLE PAGE  %%%%%%
%%%%%%%%%%%%%%%%%%%%%%%%%%%%%%%%

%  empty page follows the title page ----
\newpage
\setcounter{page}{2}
\mbox{~}
%\newpage
%
%% Author List ----------------------------
%%  You need to get a new author list!
%\input{LHCb_authorlist.tex}
%
%The author list for journal publications is provided by the Membership Committee shortly after 'approval to go to paper' has been given.
%%It will be made available on the page
%%\verb!http://www.physik.uzh.ch/~strauman/forMemCo/LHCb-PAPER-XXXX-XXX/! .
%It will be sent to you by email shortly after a paper number has beens assigned.
%The author list should be included already at first circulation, 
%to allow new members of the collaboration to verify whether they have been included correctly.
%Occasionally a misspelled name is corrected or associated institutions become full members.
%In that case, a new author list will be sent to you.
%In case line numbering doesn't work well after including the authorlist, try moving the \verb!\bigskip! after the last author to a separate line.
%
%
%The authorship for Conference Reports should be ``The LHCb
%  collaboration'', with a footnote giving the name(s) of the contact
%  author(s), but without the full list of collaboration names.

%\twocolumn
% %%%%%%%%%%%%% ---------

\renewcommand{\thefootnote}{\arabic{footnote}}
\setcounter{footnote}{0}

%%%%%%%%%%%%%%%%%%%%%%%%%%%%%%%%
%%%%%  Table of Content   %%%%%%
%%%%%%%%%%%%%%%%%%%%%%%%%%%%%%%%
%%%% Uncomment if desired
%\tableofcontents

\cleardoublepage

%%%%%%%%%%%%%%%%%%%%%%%%%
%%%%% Main text %%%%%%%%%
%%%%%%%%%%%%%%%%%%%%%%%%%

\pagestyle{plain} % restore page numbers for the main text
\setcounter{page}{1}
\pagenumbering{arabic}

%% Uncomment during review phase. 
%% Comment before a final submission.
%\linenumbers

%% This is the main body
%% It is useful to have a single file so comments are not missed in overleaf.
The study of 
charmless two-body \B-meson decays provides critical insights into strong interaction dynamics and \CP violation in heavy-quark decay processes.
 While   decays into a pseudoscalar-meson pair ($\B \to PP^{\prime} $) have been extensively examined by the \babar~\cite{BaBar:2007uoe, BaBar:2007fyq, BaBar:2007wip, BaBar:2012fgk},  \belle~\cite{Belle:2006uab, Belle:2006yry, Belle:2006uzg, Belle:2008kbm} and \lhcb~\cite{LHCb-PAPER-2014-065, LHCb-PAPER-2016-036, LHCb-PAPER-2018-006, LHCb-PAPER-2019-030, LHCb-PAPER-2020-029, LHCb-PAPER-2020-040} collaborations,  decays into a baryon-antibaryon pair are significantly less explored.  Prior to this work,  only two such processes had been observed, \BdPPbar~\cite{LHCb-PAPER-2017-022} and \BuPLbarExcited~\cite{LHCb-PAPER-2013-031},  
 and precise measurements of their decay properties are lacking.
 
 Following  pioneering searches  conducted by the \cleo~\cite{CLEO:1998wwb} and \belle~\cite{Belle:2007gob} collaborations, 
the \lhcb experiment reported the first evidence for the \BuPLbar decay with a significance of $4.1 \sigma$ using proton-proton ($pp$) collision data collected at center-of-mass energies of 7~and 8\tev during 2011--2012 (Run~1)~\cite{LHCb-PAPER-2016-048}.\footnote{The symbol $\PLambda$  is used to refer to the ground-state $\PLambda$ baryon, and the inclusion of charge-conjugate processes is implied, throughout this paper.} 
The measured  branching fraction, \mbox{$\BR(\BuPLbar) = (2.4^{+1.0}_{-0.8}\pm0.3)\times10^{-7}$}, was found to be in good agreement with the theoretical predictions~\cite{Cheng:2001tr, Chua:2013zga}, although improved experimental precision is highly desirable. This would help to better constrain the tree-level and loop (penguin) amplitudes in charmless two-body baryonic $B$-meson decays~\cite{Chua:2016aqy}, which allows for more reliable predictions of their decay rates and \CP asymmetries. Improved precision is also useful for testing
various near-threshold enhancement mechanisms~\cite{Huang:2021-review, Hsiao:2014zza, Cheng:2014qxa} proposed to explain 
why  two-body \B-meson  decays to baryons are significantly suppressed relative to their multibody counterparts.

\begin{figure}[htb]
    \begin{subfigure}{.32\textwidth}
            \centering
            \includegraphics[width=1\linewidth]{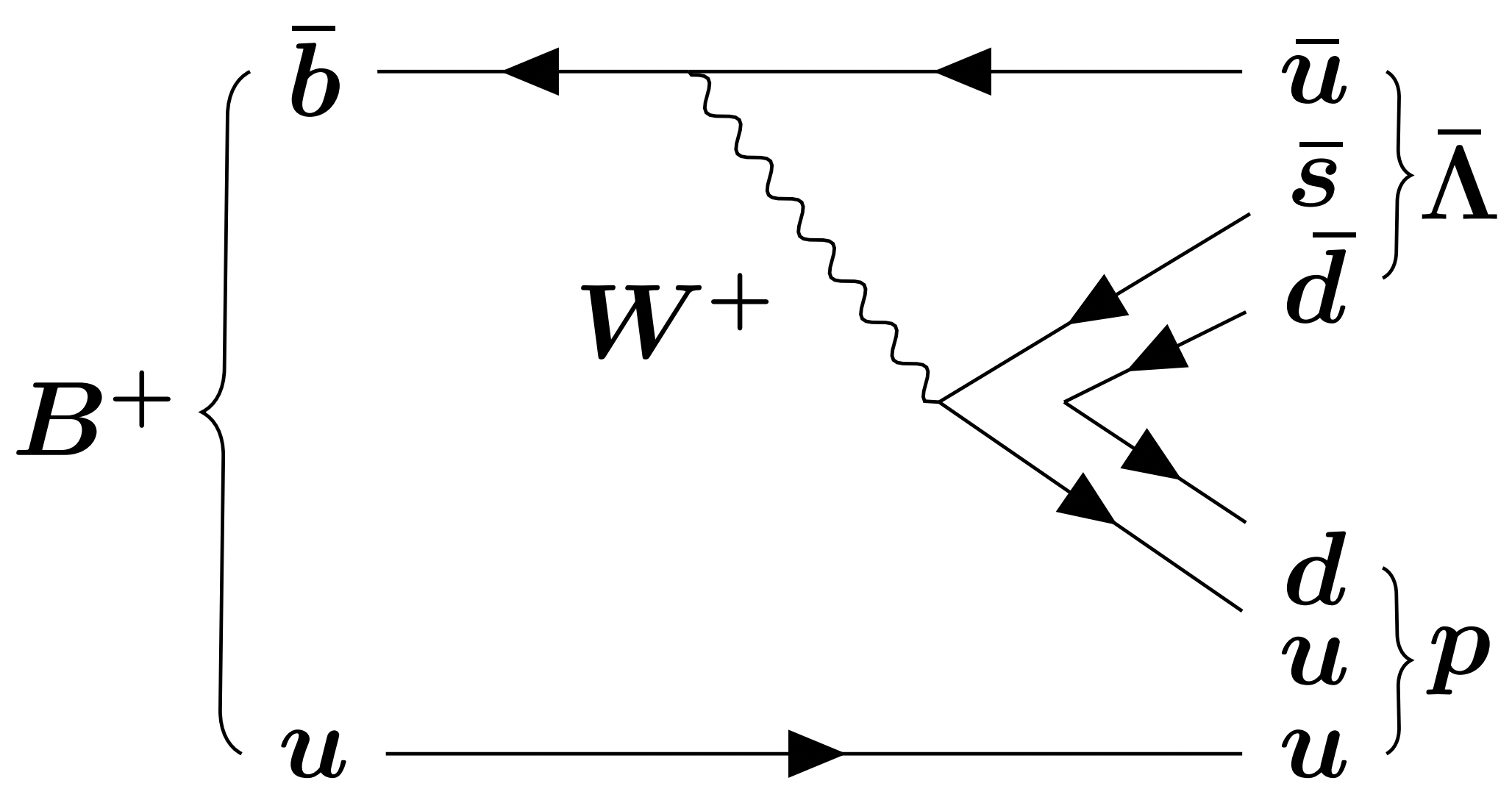}
            \caption{ Tree-level}
      \end{subfigure}
      \begin{subfigure}{.32\textwidth}
            \centering
            \includegraphics[width=1\linewidth]{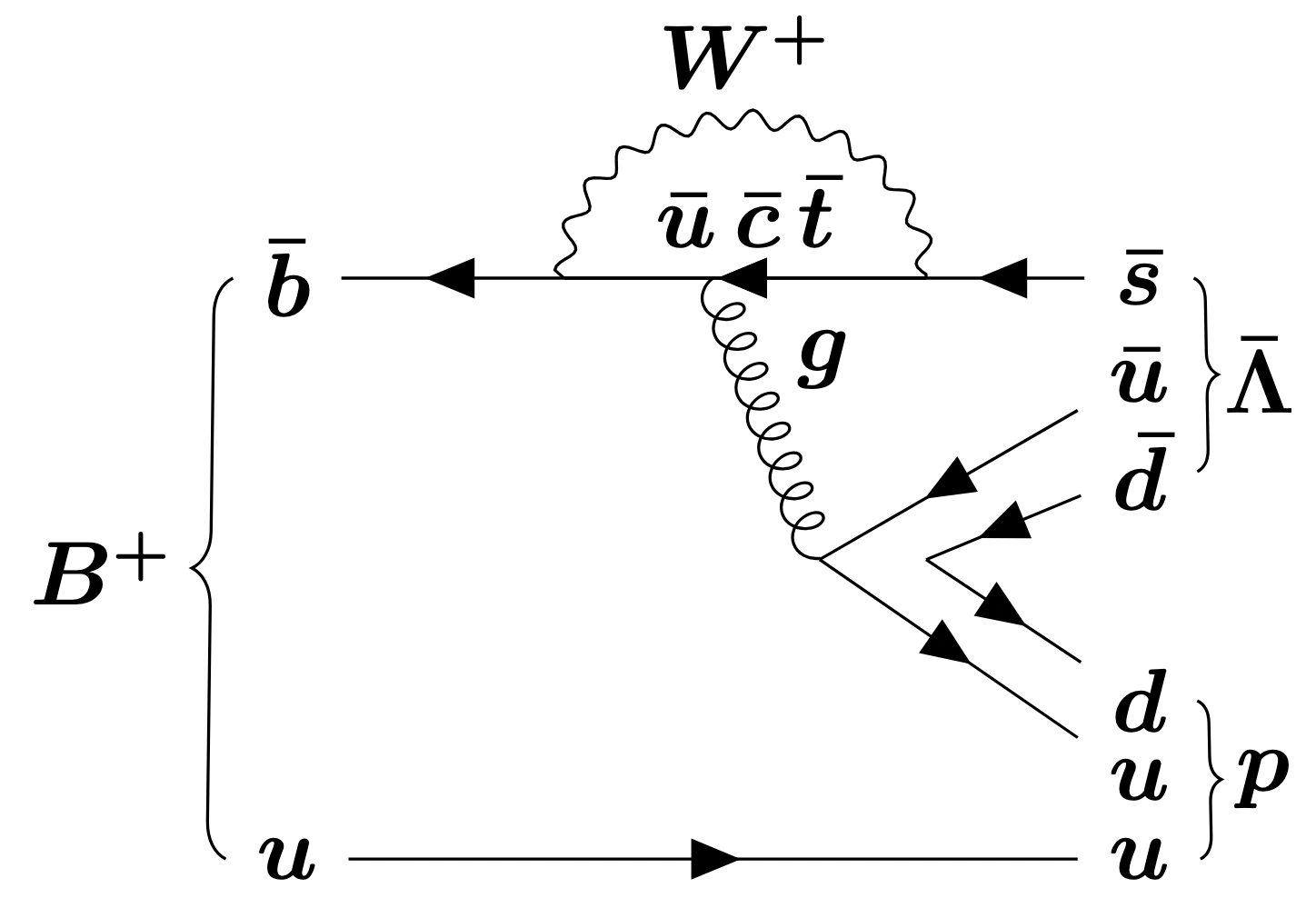}
            \caption{Penguin}
      \end{subfigure}
      \begin{subfigure}{.32\textwidth}
            \centering
            \includegraphics[width=1\linewidth]{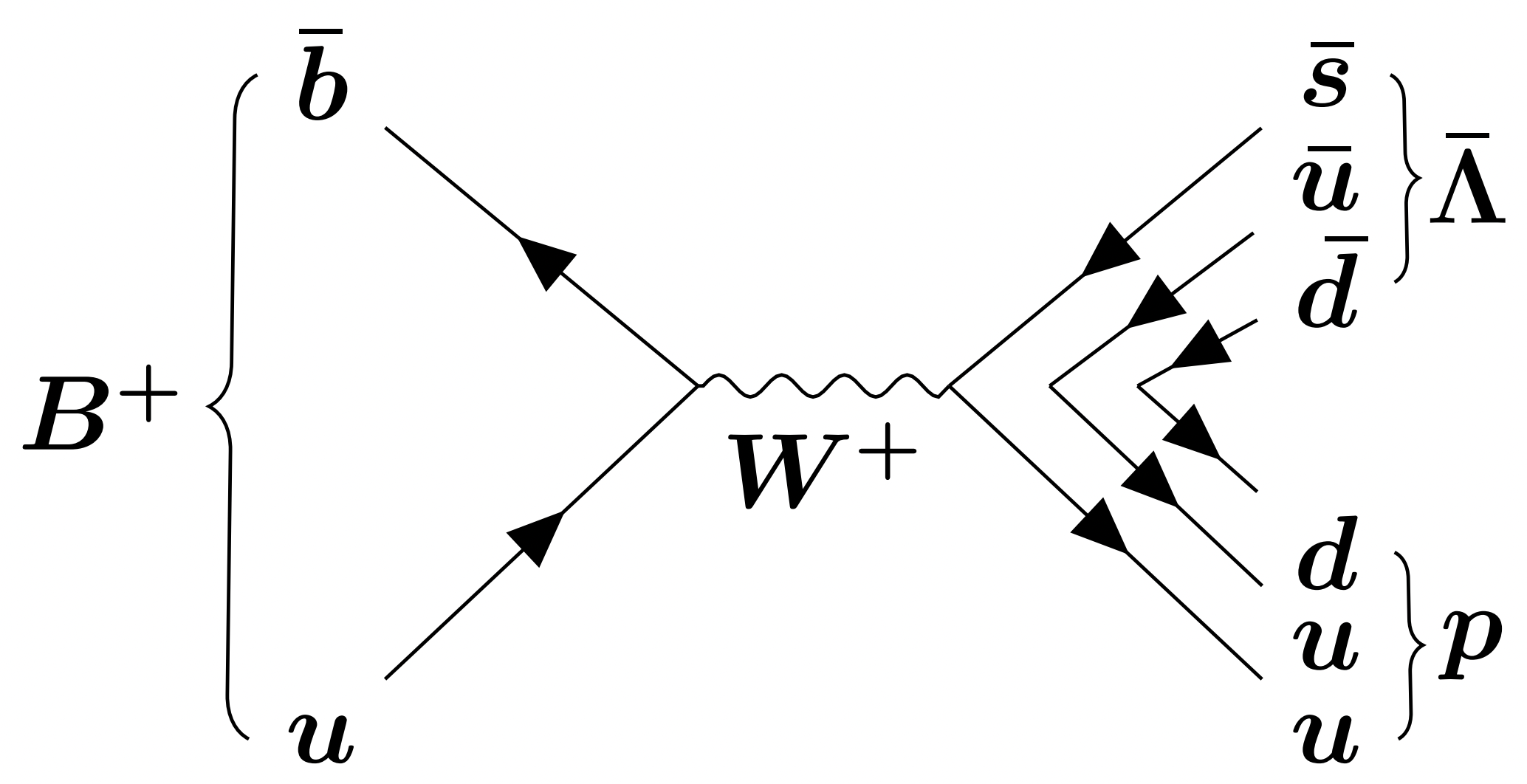}
            \caption{Annihilation}
      \end{subfigure}
    \centering
      \caption{Representative Feynman diagrams for the  $B^{+}\rightarrow p \overline{\it{\Lambda}}$  decay.}
\label{fig:feynman_diagrams}
\end{figure}

Furthermore, clearly observing the \BuPLbar decay will be a key step towards studies of \CP violation in this decay, which may potentially help resolve an intriguing anomaly regarding 
the magnitude of \CP violation observed in the \LbPK  decay.
These two decays, as well as the  \BsKK  and \BdKPi  decays, are related by the underlying  \btouus quark transition. The representative diagrams for  \BuPLbar are illustrated in Fig.~\ref{fig:feynman_diagrams}.
Sizable \CP-violation effects can arise from the interference between tree-level and penguin contributions of similar size~\cite{Han:2024kgz}.
While the \BsKK and \BdKPi   decays exhibit \CP asymmetries of $\sim 10\%$~\cite{LHCb-PAPER-2020-029}, the recent  measurement of \mbox{$A_{\CP}(\LbPK ) =(-1.14\pm 0.76)\%$}~\cite{LHCb-PAPER-2024-048} by the \lhcb collaboration reveals a puzzling suppression.
This anomaly may stem from cancellation of \CP asymmetries between different partial waves~\cite{Han:2024kgz, Han:2025tvc}  --- a unique feature present in $b$-hadron decays involving baryons, but absent in $\B \to PP^{\prime} $ processes.
Theoretical studies have 
suggested that the  \CP asymmetry in \BuPLbar could reach the $10\%$ level~\cite{Chua:2016aqy}, but this could be suppressed by cancellation between the S-wave and P-wave amplitudes~\cite{Geng:2023nia}, $A_{\mathrm{S}}$ and $A_{\mathrm{P}}$, corresponding  to the decays to states with orbital angular momentum  $L=0$ and $L=1$ between the proton and the $\Lbar$ baryon.
However, this cancellation necessitates both sizable \mbox{S-wave} and P-wave amplitudes.
An angular analysis measuring the weak decay parameter \mbox{$\alpha_B=2\Real(A_{\mathrm{S}} \cdot A_{\mathrm{P}})/(|A_{\mathrm{S}}|^2 + |A_{\mathrm{P}}|^2)$}~\cite{Han:2024kgz},  which characterizes  the interference between the two amplitudes, can test this assumption.
Therefore, experimental investigation of the \CP violation 
and angular distribution in 
\BuPLbar decays 
can provide valuable insights  
to elucidate the situation.

This Letter presents the first observation of the rare baryonic decay  \BuPLbar
and the first measurement of its decay parameter
using $pp$ collision data collected during 2016--2018 (Run~2) by the \lhcb experiment at a center-of-mass energy of 13\tev, 
corresponding to an integrated luminosity of 5.4\invfb. 
The branching fraction of the \BuPLbar decay was measured relative to the  normalization decay  \BuKSPi, with subsequent decays \LbarPPi  and \KSPiPi.
This result was then combined with the previous \lhcb measurement.
In order to avoid experimenter’s
 bias, the results of the analysis were not examined until the full analysis procedure had been finalized.

The \lhcb detector is a single-arm forward spectrometer covering the \mbox{pseudorapidity} range $2<\eta <5$, described in detail in Refs.~\cite{LHCb-DP-2008-001,LHCb-DP-2014-002}. 
Of particular relevance for this analysis is the tracking system.
Depending on the decay vertex position of the   
\Vz (representing \Lbar or $\KS$) particle, the decay products can be reconstructed as {\it long} tracks using hits from the full tracking system, or as {\it downstream} tracks  without  hits in the vertex detector.  
This analysis uses \Vz decays reconstructed 
from either two long tracks (LL) or two downstream tracks (DD). Due to their difference in the  invariant-mass resolution, 
\Bu candidates formed with these two types of \Vz hadrons are analyzed separately. Particle identification (PID) information is provided by two ring-imaging Cherenkov detectors.

Simulation is used to develop the selection strategy, model the signal and background invariant-mass distributions,  and evaluate the efficiencies for the measurement of the relative branching fraction. Signal and normalization decays are
generated uniformly in phase space
using the software described in Refs.~\cite{Sjostrand:2007gs,*Sjostrand:2006za,Lange:2001uf,Golonka:2005pn,Allison:2006ve, *Agostinelli:2002hh,LHCb-PROC-2011-006}.
The simulated kinematic spectra of the \Bu mesons, angular distribution of the final-state particles,  as well as the track multiplicity, are corrected to match those observed in the data.
The PID response in the simulation is also tuned to match that in the data using dedicated calibration samples\cite{LHCb-DP-2018-001}.

The online event selection~\cite{LHCb-DP-2019-001} is performed by a trigger~\cite{LHCb_trigger}, which consists of a hardware stage, based on information from the calorimeter and muon systems, followed by two software stages, which apply a partial and full event reconstruction. The hardware trigger decisions are classified as trigger on signal (TOS) if only signal tracks are used in the trigger decision, or as trigger independent of signal (TIS) if particles independent of the signal, such as high energy leptons or photons from the decay of the other $b$-hadron, are used to select the event.
Subsequently, the software trigger requires a two- or three-track secondary vertex with a significant displacement from any primary $pp$ interaction vertex (PV).

In the offline selection, 
four independent subsamples of \Bu candidates  are formed for each decay mode, corresponding to the combinations of two trigger categories (TOS and TIS) and two track categories (LL and DD).
Due to their different  background levels and compositions, these categories  are analyzed separately in the selection and mass fit procedures, but they are combined for the angular analysis.

The \Vz candidates, reconstructed using pairs of oppositely charged tracks under the \Pbarpi  or  \PiPi  particle hypothesis, must have a good-quality decay vertex that is significantly displaced from any PV and meet minimum momentum thresholds.
The reconstructed invariant mass is required to fall in the window $[1112, 1120]\mevcc$ ($[1110, 1122]\mevcc$) for the LL (DD) category under the \Pbarpi  hypothesis for $\Lbar$ candidates, and $[480, 515]\mevcc$ ($[465, 530]\mevcc$) under the $\pip\pim$ hypothesis for \KS candidates.
Subsequently, the \Vz candidates are further combined with 
a proton (pion)  for the signal (normalization) channel to reconstruct \Bu candidates. 
The \Bu and \Vz decay vertices are required to be separated by at least  2.5 standard deviations along the beam direction.
Since the momentum vector of a \Bu candidate should point back to the production vertex, the candidate is required to have
 a small opening angle between its  momentum  and  flight direction,
and  a small \chisqip  with respect to the associated PV. 
Here the associated PV  is the PV that is best aligned with the \Bu flight direction, and \chisqip is the difference between the vertex-fit \chisq of the given PV reconstructed with and without the particle in question.
A kinematic fit is performed to improve the \Bu  mass resolution by constraining the invariant mass of the \Vz candidate to the known  $\Lbar$ (\KS) mass\cite{PDG2024}
and requiring the \Bu candidate to originate from the associated PV.

Due to possible misidentification of a charged pion as a proton, or vice versa, \KSPiPi and \Psipp (only in the LL case) decays can be  mistakenly reconstructed as  \LbarPPi decays. These background candidates are suppressed using PID information or rejected by 
dedicated vetoes in the corresponding mass spectra. 
Similarly, for the \BuKSPi mode, backgrounds from \mbox{\KStarKpi} (LL only), \mbox{\LbarPPi}, and \DKpi (LL only) arise when a charged kaon or proton is misidentified as a pion, and these candidates are vetoed.

For each \Bu decay mode and data category, a multivariate classifier based on the boosted decision tree~(BDT)~\cite{Breiman, AdaBoost} method
is employed to differentiate  the signal from combinatorial
background composed of random combinations of tracks.
 The classifier is trained using simulated signal decays and  background  from the high invariant-mass sideband ([5700, 6000]\mevcc) of selected \Bu data candidates.
The BDT classifier mainly uses information on the decay kinematics and topology of the \Bu decay chain, as well as PID information for the charged hadron originating directly from the \Bu decay.
The optimal requirement on the BDT response 
of the  \BuPLbar candidates is determined  by maximizing the figure-of-merit $\epsilon/(\frac{\mathrm{X}}{2}+\sqrt{N_{B}})$~\cite{Punzi:2003bu}, where $\epsilon$ is the signal efficiency of the BDT requirement obtained from simulation, $\mathrm{X}$ is the target significance (set to 5), and $N_{B}$ is the background yield in the  invariant-mass signal region defined to be within three times the resolution from the known \Bu mass~\cite{PDG2024}.
The BDT requirement for the \BuKSPi mode is optimized by maximizing the quantity $N_{S}/\sqrt{N_{S}+N_{B}}$,
where $N_{S}$ denotes the expected signal yield.
After applying all the selection criteria discussed above, 
all candidates in the \Bu invariant-mass region \mbox{[5040, 5700]\mevcc} are retained for the subsequent analysis.

To determine the signal yields, a simultaneous unbinned maximum-likelihood fit is performed for each data category on both the \PLbar and \KSPi mass distributions. The \PLbar mass spectrum consists of a signal peak,  a combinatorial background component, and  
partially reconstructed structures appearing in the low-mass region. 
The \KSPi mass spectrum features an additional misidentification background from $\BuKSK$ decays.

For both decay modes, the signal peak is described by the sum of a double-sided crystal-ball (DSCB) function~\cite{Skwarnicki:1986xj} and a Gaussian function. The tail parameters, the relative fractions of the two functions and the ratio of their widths are fixed to values obtained from simulation.  
The combinatorial background is described by an exponential function, with the slope parameter determined from the fit to each mass spectrum. 
Partially reconstructed backgrounds populating the \BuPLbar lower-mass sideband are assumed to arise from \BdPLbarPi, \BuPLbarPiz and \BuPLbarGamma decays. 
Their shapes are derived from simulation and modelled using a kernel density estimation (KDE)~\cite{rookeyspdf198}.
%Their shapes are derived from simulation using a nonparametric probability density function obtained via a kernel density estimation procedure~\cite{rookeyspdf198}. 
In the \KSPi mass spectrum, the  partially reconstructed 
background is modeled empirically by an \textsc{ARGUS} function~\cite{ARGUS:1990hfq} convolved with a Gaussian function.
Furthermore, the background due to misidentified  $\BuKSK$ decay is modeled by a DSCB function, with the tail parameters and the mass difference between the $\BuKSK$ and $\BuKSPi$ peaks fixed to values obtained from simulation.

In the simultaneous fit, the \BuPLbar and \BuKSPi signal models share a common peak position and a  scale factor that describes potential differences in mass resolution between data and simulation. 
The yields of all components are allowed to vary in the fit, except the yield of the misidentification background from \BuKSK decays, which is fixed based on the $\BuKSK$ to $\BuKSPi$ branching fraction ratio and the relative efficiencies determined from simulation.

\begin{figure}[t]
    \centering
    \includegraphics[width=0.48 \textwidth]{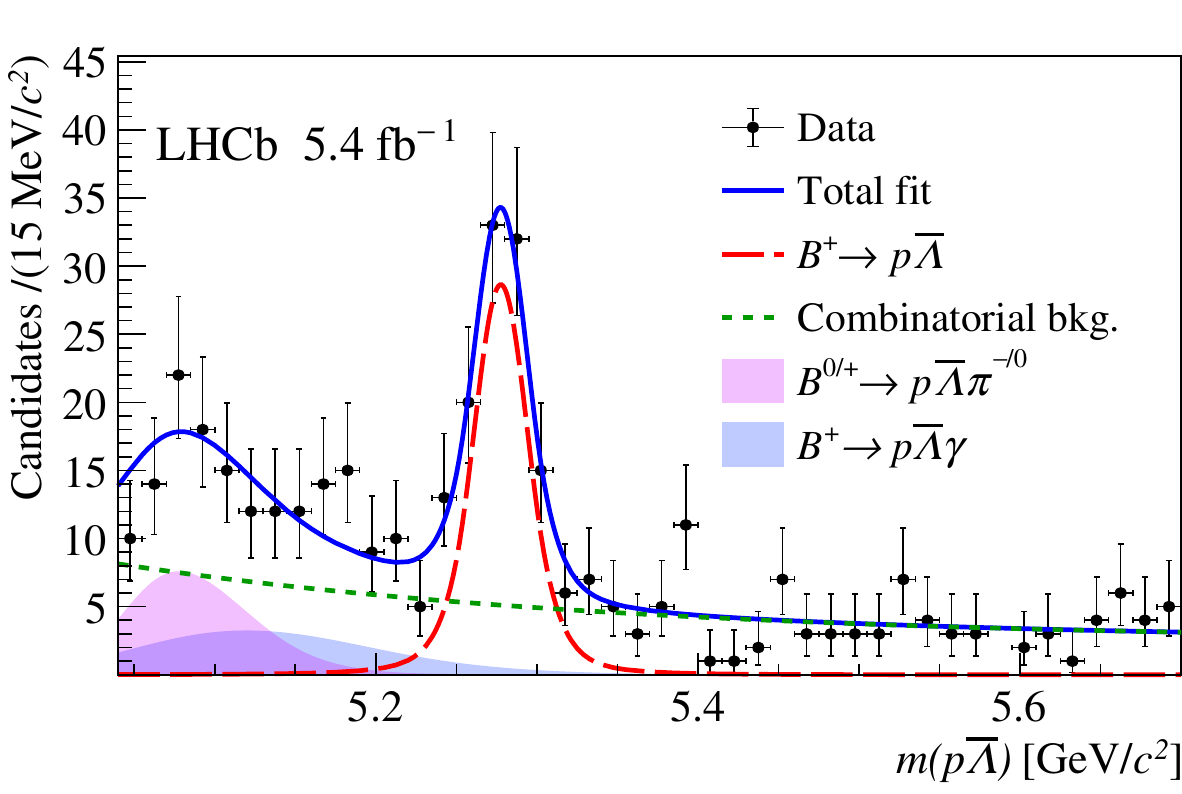}
    \includegraphics[width=0.48 \textwidth]{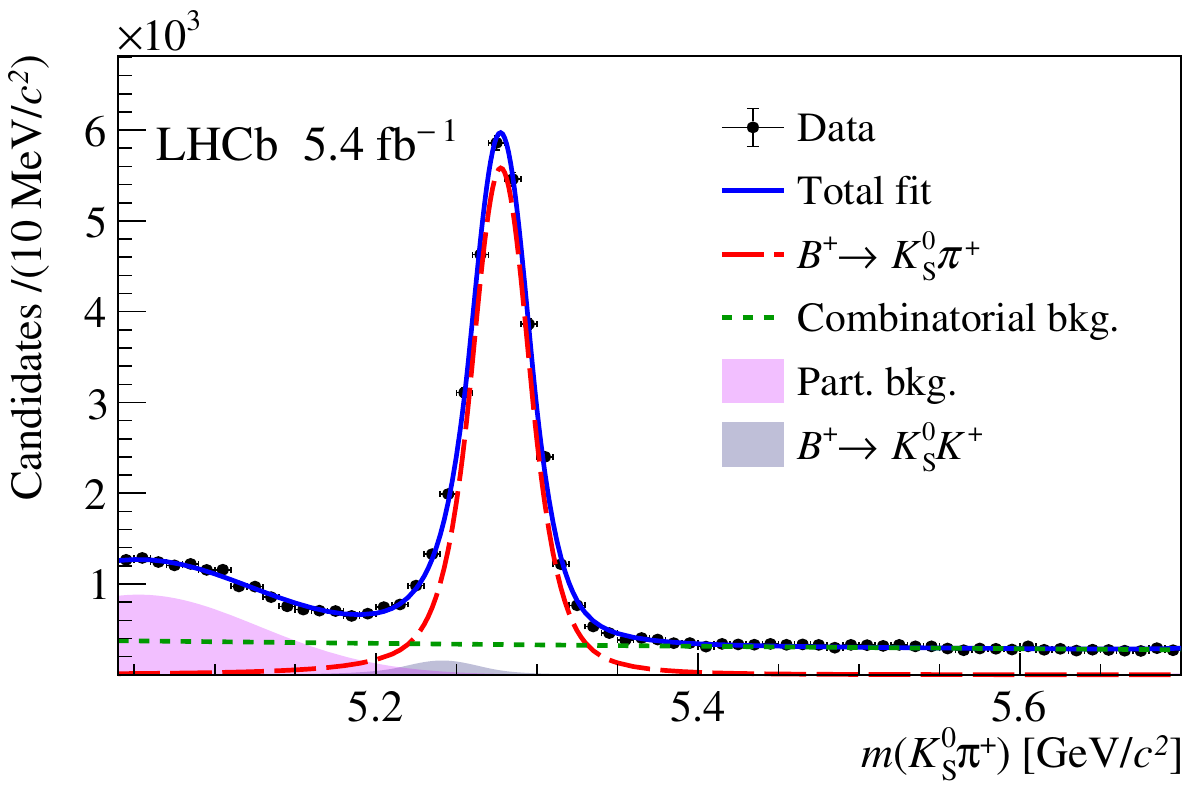}
    \caption{Invariant-mass distributions of selected (left) $B^{+}\rightarrow p \overline{\it{\Lambda}}$ and  (right) $B^{+} \rightarrow K^{\mathrm{0}}_{\mathrm{S}} \pi^{+}$ candidates, combining  all categories.
    The results of the fit are also shown.
    }
    \label{fig:Massfit_B2pL_B2Kspi_data}
\end{figure}

The obtained yields of 
\BuPLbar and \BuKSPi  decays 
in each  data category
are  given in  the End Matter.
The total signal yield is determined to be $88\pm12$ for the $\BuPLbar$ decay
and $(28.5\pm0.2)\times10^{3}$  for 
the  $\BuKSPi$ process,   where the uncertainties are statistical only. 
The invariant-mass distributions for both \Bu decay modes,  summing over all four data categories,
%of the four categories 
are shown in Fig.~\ref{fig:Massfit_B2pL_B2Kspi_data} along with the fit results. 
The statistical significance of the $\BuPLbar$ decay, evaluated using the likelihood-ratio test~\cite{Wilks:1938dza}
and combining the four subsamples, exceeds seven standard deviations. The log-likelihood curve is shown in the End Matter. The  fit
results are used to subtract the background in the subsequent angular analysis, using the \splot technique~\cite{Pivk:2004ty}.
The outcome of the angular analysis is in turn used to determine the decay branching fraction, as described later.

\begin{figure}[t]
    \centering
    \includegraphics[width=0.48 \textwidth]{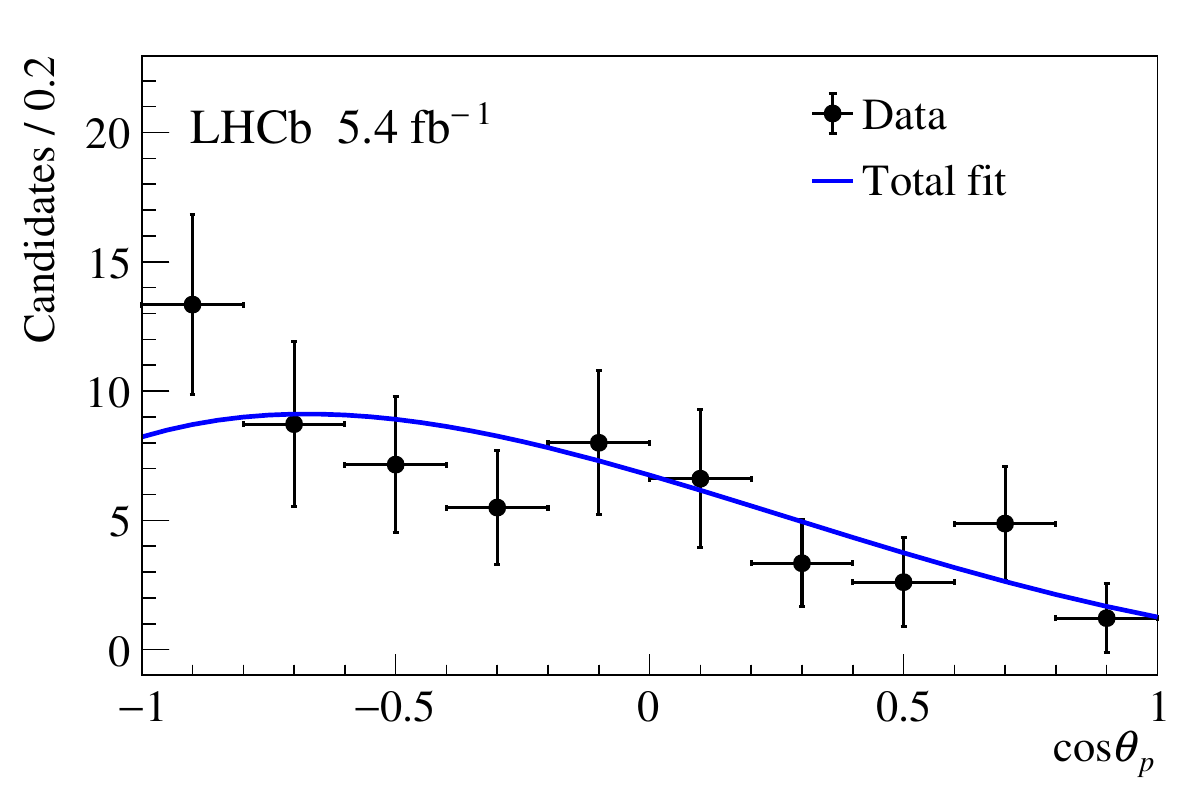}
    \includegraphics[width=0.48 \textwidth]{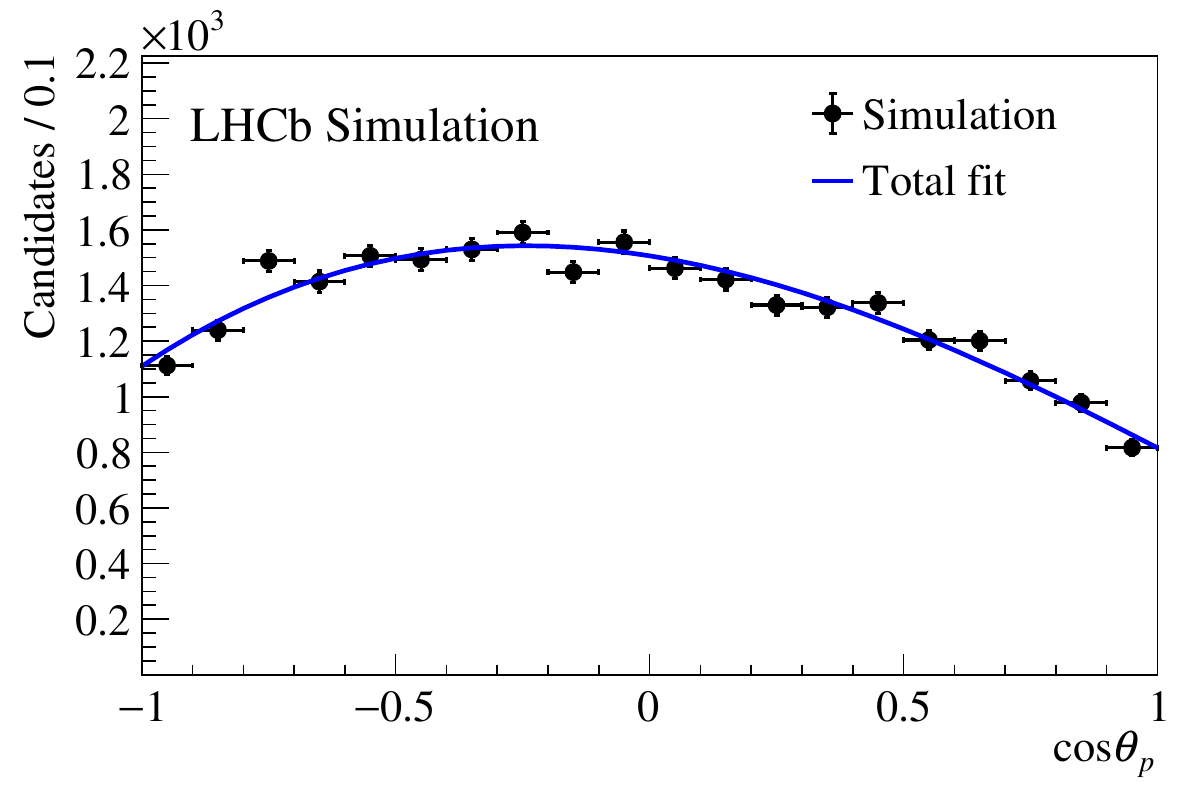}
    \caption{Distributions of $cos(\theta_{p})$ in  (left)  the background-subtracted $B^{+}\rightarrow p \overline{\it{\Lambda}}$ data  and (right) the simulated $B^{+}\rightarrow p \overline{\it{\Lambda}}$  sample generated uniformly in phase space, together with the fit results. 
     }
    \label{fig:fig_B2pL_anglefit}
\end{figure}

The weak decay parameter \alphaB can be obtained from an analysis of the angular distribution described by~\cite{Jacob:1959at} 
\begin{align}
\label{eq:angular_dis}
    \frac{\deriv N}{\deriv\costhetap} \propto (1-\alphaL\alphaB\costhetap) \cdot  \epsilon(\costhetap), \nonumber
\end{align}
where $\thetap$ denotes 
the angle between the antiproton momentum and the direction opposite to the $B^{+}$ momentum in the $\Lbar$ rest frame,
 and \alphaL is the weak decay parameter of the  \LPPi  process, for which the \CP-average value $\alpha^{\rm avg}_{\Lambda}=0.754\pm0.003$ reported by the  BESIII experiment is used~\cite{BESIII:2022qax}. The acceptance function 
 $\epsilon(\costhetap)$  accounts for the effect of the  detector geometry and selection requirements. Its shape  is obtained from simulation
 and described using a second-order polynomial function. 
 The background-subtracted angular distribution  and the acceptance shape are presented in Fig.~\ref{fig:fig_B2pL_anglefit}.
 An unbinned  maximum-likelihood fit to the  background-subtracted  data~\cite{Xie:2009rka} is performed to extract  the decay parameter $\alphaB$.
The angular fit procedure is validated using pseudoexperiments, 
and a small bias in the estimate of \alphaB is observed and assigned as a systematic uncertainty.
 The evaluation of the acceptance shape from simulation is verified using the $\BuKSPi$ control channel, for which the distribution of the cosine of the $\pip$ helicity angle in the \KS rest frame is known to be flat.  Systematic uncertainties associated with  the modeling of  the acceptance shape and the signal and background mass shapes are assessed by using alternative models, while biases due to the assumption of no correlation between the $\PLbar$ mass and the angular variable $\costhetap$ are evaluated  by performing separate mass fits in different ranges of $\costhetap$ and recalculating the signal weights. The resulting shifts of the $\alphaB$ value are assigned as the corresponding systematic uncertainties and summed in quadrature. The uncertainty originating from the $\PLambda$ weak decay parameter is found to be negligible.
 The decay parameter  is measured to be 
 \begin{equation*}
 \alpha_B = 0.87_{-0.29}^{+0.26} \pm 0.09, 
 \end{equation*}
 where the first uncertainty is statistical and  the second is systematic.
 This result is consistent with the theoretical prediction of $\alphaB \approx 0.6$~\cite{Geng:2023nia}.

The branching fraction of the $\BuPLbar$ decay is measured  with respect to that of the $\BuKSPi$ decay
following the equation 
\begin{align}
\label{eq:BFRatio}
    \mathcal{R} \equiv \frac{\mathcal{B}(\BuPLbar)}{\mathcal{B}( \BuKSPi)}
    &=\frac{N(\BuPLbar)}{N(\BuKSPi)} \cdot \frac{\epsilon(\BuKSPi)}{\epsilon(\BuPLbar)} \cdot \frac{  \mathcal{B}( \KSPiPi ) }{  \mathcal{B}(\LbarPPi)  } . \nonumber
\end{align}
Here \mbox{$\mathcal{B}( \KSPiPi  ) = (69.20\pm0.05) \times 10^{-2}$} and \mbox{$\mathcal{B}(\LPPi ) = (64.10\pm0.50) \times 10^{-2}$} are taken from Ref.~\cite{PDG2024}, $N$ denotes the  yield of the signal or normalization channel extracted from the data fits, and $\epsilon$ represents the total reconstruction and selection efficiency, which is determined for each decay mode using corrected simulation samples.

Systematic uncertainties on the branching fraction ratio
associated with the fitting procedure and the efficiency determination are  summarized in Table~\ref{tab:Allrelativesystematics:summary}.
Pseudoexperiments are performed to validate the mass fit procedure. A bias in the \BuPLbar yield is found and assigned as a systematic uncertainty.
The systematic uncertainty resulting from the choice of fit model is quantified using alternative models. As examples, the DSCB function
for the signal peak is replaced by a Hypatia function~\cite{Santos:2013gra}, and a second-order Bernstein polynomial is used instead of an exponential function for the combinatorial background.
Uncertainties associated with the remaining fit components are found to be negligible.

The statistical uncertainties of the efficiencies due to the finite size of the simulation samples are considered as a source of systematic uncertainty and propagated to the branching fraction ratio. 
The simulation-based efficiency estimation relies on accurately associating generated and reconstructed particles. The effect of imperfect matching is investigated by redetermining the efficiencies with an alternative method.
Furthermore, the hardware-trigger efficiency is calibrated using a data-driven method~\cite{MartinSanchez:1407893}. 

\begin{table}[tb]
\caption{Summary of relative systematic uncertainties (in percent) for the measured ratio of branching fractions in the four categories. The total relative uncertainties, obtained by summing all contributions in quadrature, are provided for comparison.}
\centering
\small
\begin{tabular}{l S[table-format=2.1] S[table-format=2.1] S[table-format=2.1] S[table-format=2.1]} 
\hline
Source              & {TOS LL} & {TOS DD} & {TIS LL} & {TIS DD} \\
\hline
      Fit bias      &0.3  &1.0   &2.1  &0.9   \\
      Fit model              &2.9  &4.6   &4.0 &12.4   \\
      Simulation sample size &1.1  &1.2  &1.8  &3.1   \\
      Data-simulation matching    &0.6  &1.0  &3.3  &4.8   \\
   Trigger efficiency   &2.2  &2.2  &1.4  &1.9   \\
   Tracking efficiency     &1.0  &1.3  &1.0  &1.4   \\
    Kinematic correction     &3.5  &2.2  &0.0  &0.5   \\
          PID                &0.3  &0.9  &0.5  &3.5   \\
        \Lbar polarization   &1.0  &0.9  &2.1  &2.2   \\
\hline
    $K^{0}_{S}/\PLambda$ branching fraction ratio &\multicolumn{4}{c}{1.0} \\
\hline
      Total                  &5.4  &6.2  &6.4  &14.3   \\
\hline
      
\hline
\end{tabular}
\label{tab:Allrelativesystematics:summary} 
\end{table}

Systematic uncertainties associated with the corrections applied to simulation are estimated by using alternative correction schemes or control samples and quantifying the resulting variations in efficiency. 
The efficiencies  could depend on the  polarization of the  \Lbar baryon, which is set to the measured  $\alphaB$ value for the efficiency estimation.
Uncertainties related to the \Lbar polarization are evaluated by varying the  $\alphaB$ value up and down by one standard deviation and calculating the efficiency shifts.

 Among the systematic uncertainties described  above, the uncertainties associated with the trigger, tracking, PID efficiency corrections and \Lbar polarization are considered fully correlated across the four data  categories due to their reliance on shared calibration sources. All other uncertainties are treated as uncorrelated between different categories. These uncertainties, along with those on the branching fractions of the \LbarPPi and \KSPiPi decays, are added in quadrature to give the total systematic uncertainty on the branching fraction ratio $\mathcal{R}$.

Taking into account the  correlations of the systematic uncertainties  between different data categories, the average result from LHCb Run~2 data is determined to be \mbox{$\mathcal{R}^{\rm Run2}= (1.01\pm0.14\pm0.04)\times10^{-2}$},
which is \mbox{compatible} with the LHCb Run~1 result \mbox{$\mathcal{R}^{\rm Run1}= (2.02\pm0.88\pm0.23)\times10^{-2}$}~\cite{LHCb-PAPER-2016-048}. 
Neglecting the small correlations between the Run~1 and Run~2 results, the combined result is
\begin{equation}
\mathcal{R}^{\rm Run1\&2}= (1.04\pm0.14\pm0.04)\times10^{-2}\,.\nonumber
\end{equation}
 Using the known branching fraction $\mathcal{B}(\BuKSPi)= (1.19\pm0.03)\times10^{-5}$~\cite{PDG2024}, the absolute branching fraction is determined to be
\begin{equation}
    \BF(\BuPLbar)  = (1.24\pm0.17\pm0.05\pm0.03)\times10^{-7}\,,\nonumber
\end{equation}
where 
the third uncertainty is due to knowledge of the $\BuKSPi$ branching fraction.

In summary, 
a search for  the rare  decay \BuPLbar is performed using a sample of \proton\proton collision data collected at a center-of-mass energy of 13\tev, corresponding to an integrated luminosity of 5.4\invfb. 
An excess of signal candidates over the expected background is observed with a statistical significance exceeding seven standard deviations, when accounting for systematic uncertainties. This represents the first observation of this decay.
The measured branching fraction is consistent with available
theoretical predictions~\cite{Cheng:2001tr, Chua:2013zga} and is more than twenty times lower than that of the \BuPLbarPiz  mode~\cite{PDG2024}, supporting the theoretical expectation of threshold enhancement in  three-body baryonic \B decays. The large value of the measured weak decay parameter $\alphaB$ indicates a strong interference between the competing S-wave and P-wave amplitudes of the \BuPLbar decay, which may lead to substantial cancellation between the \CP asymmetries of different partial waves.

It is noteworthy that the branching fraction of the  \BuPLbar decay, proceeding via a \btouus  transition,  is an order of magnitude higher than that of the \BdPPbar  decay~\cite{LHCb-PAPER-2022-004}, which proceeds via  a \btouud  transition.  This pattern mirrors that in mesonic two-body $B$-meson decays, where  \btouus transitions (\eg, \BsKK  and \BdKPi )
dominate over  \btouud transitions (\eg, \BsPiK  and  \BdPiPi ). However, this 
contrasts with the situation in $b$-baryon decays, where
these two types of processes (\eg, 
\LbPK and \LbPPi ) exhibit similar branching fractions~\cite{PDG2024}. 
These  distinctive patterns, in conjunction with the discrepancy in the measured \CP violation in \BsKK , \BdKPi   and \LbPK  decays, motivate future precision studies.  
Leveraging the large datasets anticipated from the upgraded \lhcb  experiment in future running periods, \CP violation measurements in the \BuPLbar decay channel will become feasible, offering a powerful probe into the rich dynamics of heavy-flavor decays and the mechanisms of \CP violation.

%\clearpage 

% Do not include this in any draft (just for information in the template)

%\input{acknowledgements_intro}
% Comment this in for paper drafts; do not include this in analysis note, conference and figure reports
\emph{Data availability:} The data that support the findings of this article are openly available~\cite{LHCb-PAPER-2025-044-cds}.
\section*{Acknowledgements}
%
% These Acknowledgements valid from 12/09/25
%
\noindent We express our gratitude to our colleagues in the CERN
accelerator departments for the excellent performance of the LHC. We
thank the technical and administrative staff at the LHCb
institutes.
We acknowledge support from CERN and from the national agencies:
ARC (Australia);
CAPES, CNPq, FAPERJ and FINEP (Brazil); 
MOST and NSFC (China); 
CNRS/IN2P3 (France); 
BMFTR, DFG and MPG (Germany);
INFN (Italy); 
NWO (Netherlands); 
MNiSW and NCN (Poland); 
MCID/IFA (Romania); 
%MSHE (Russia); 
MICIU and AEI (Spain);
SNSF and SER (Switzerland); 
NASU (Ukraine); 
STFC (United Kingdom); 
DOE NP and NSF (USA).
%%%%%%%%%%%%%%%%%%%%%%%%%%%%%%%%%%%%%%%%%%%%%
We acknowledge the computing resources that are provided by ARDC (Australia), 
CBPF (Brazil),
CERN, 
IHEP and LZU (China),
IN2P3 (France), 
KIT and DESY (Germany), 
INFN (Italy), 
SURF (Netherlands),
Polish WLCG (Poland),
IFIN-HH (Romania), 
%RRCKI and Yandex LLC (Russia), 
PIC (Spain), CSCS (Switzerland), 
and GridPP (United Kingdom).
%%%%%%%%%%%%%%%%%%%%%%%%%%%%%%%%%%%%%%%%%%
We are indebted to the communities behind the multiple open-source
software packages on which we depend.
%%%%%%%%%%%%%%%%%%%%%%%%%%%%%%%%%%%%%%%%%%
Individual groups or members have received support from
%ARC and ARDC (Australia); % moved to national 16/01
Key Research Program of Frontier Sciences of CAS, CAS PIFI, CAS CCEPP, 
%Fundamental Research Funds for the Central Universities,  and Sci.\
%\& Tech.\ Program of Guangzhou (China); Removed 24/11/25
Minciencias (Colombia);
EPLANET, Marie Sk\l{}odowska-Curie Actions, ERC and NextGenerationEU (European Union);
A*MIDEX, ANR, IPhU and Labex P2IO, and R\'{e}gion Auvergne-Rh\^{o}ne-Alpes (France);
%RFBR, RSF and Yandex LLC (Russia);
Alexander-von-Humboldt Foundation (Germany);
ICSC (Italy); 
%GVA, XuntaGal, GENCAT, Inditex, InTalent and Prog.~Atracci\'on Talento, CM (Spain);
Severo Ochoa and Mar\'ia de Maeztu Units of Excellence, GVA, XuntaGal, GENCAT, InTalent-Inditex and Prog.~Atracci\'on Talento CM (Spain);
SRC (Sweden);
the Leverhulme Trust, the Royal Society and UKRI (United Kingdom).

\clearpage

\section*{End Matter}
\setcounter{section}{0}

\section{Additional mass fit and angular fit results}

\begin{figure}[hbt]
     \centering
     \includegraphics[width=0.45 \textwidth]{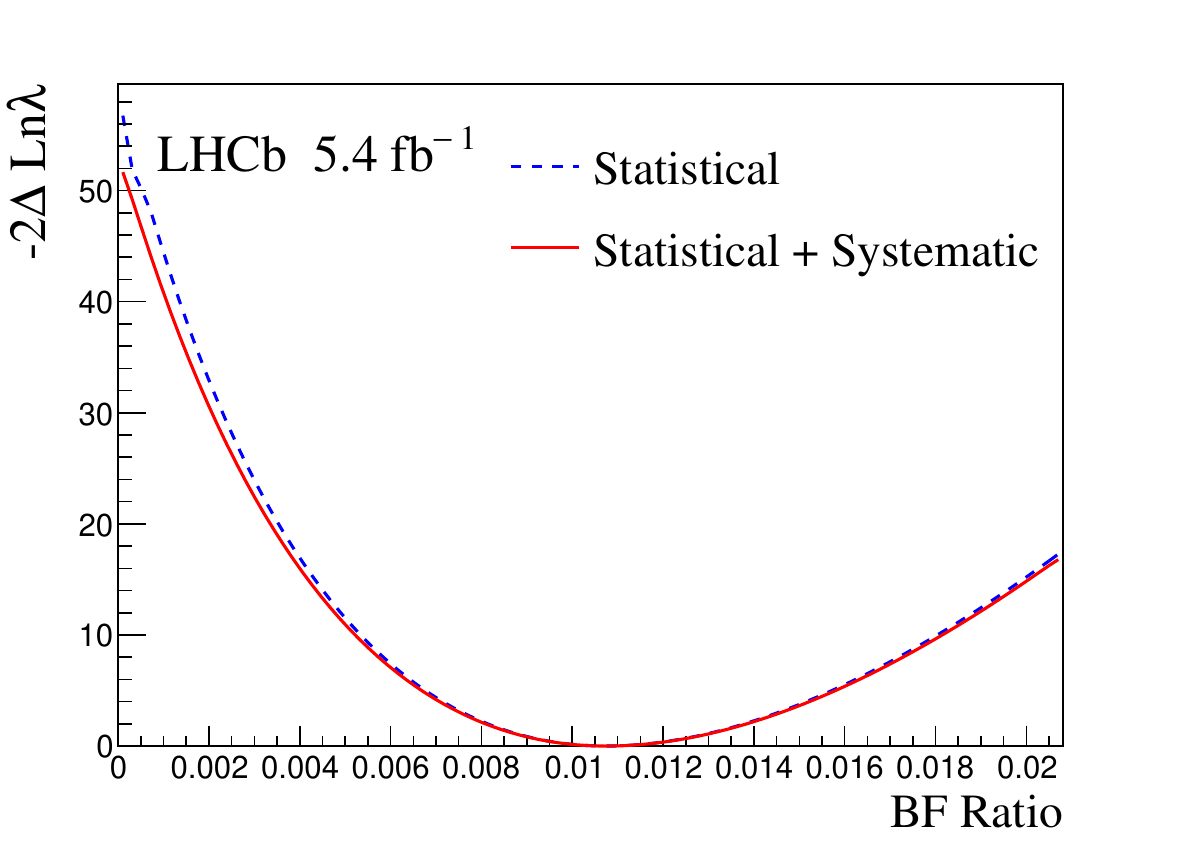}
     \includegraphics[width=0.45 \textwidth]{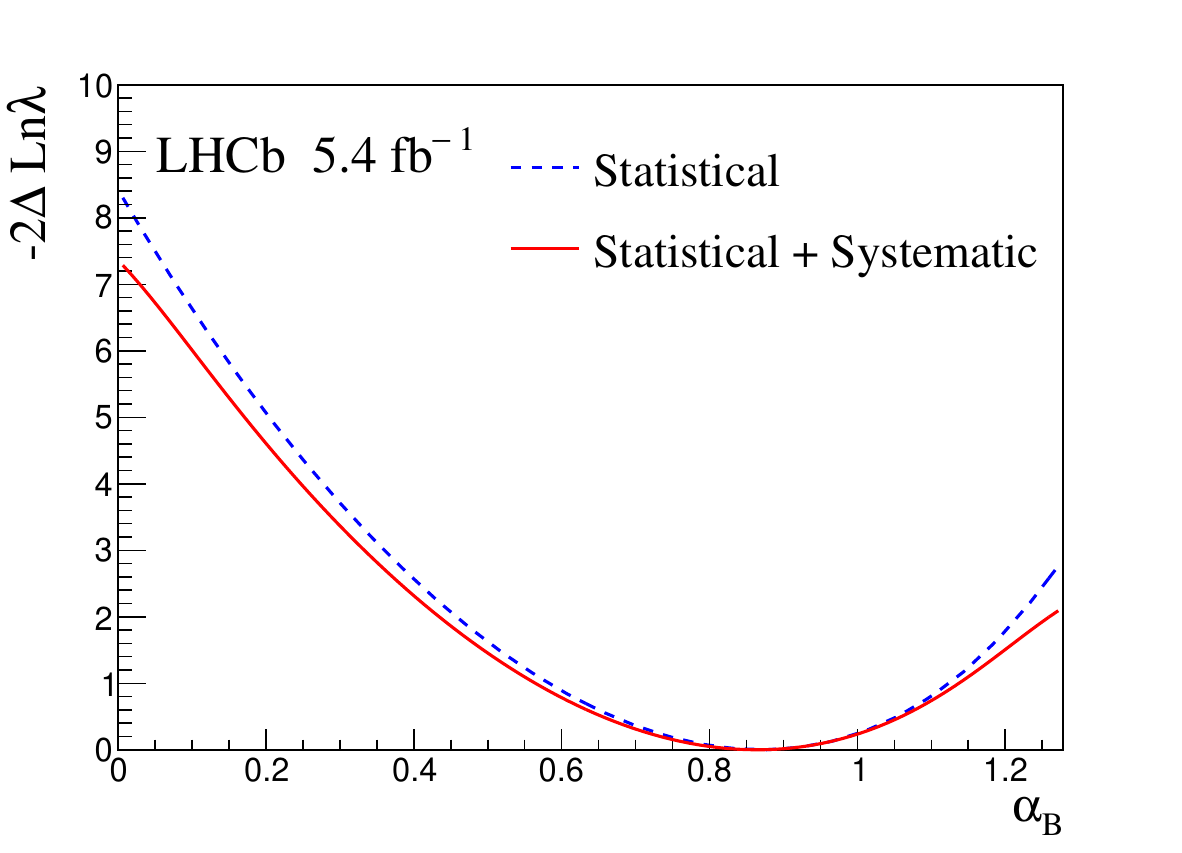}
     \caption{Profile-likelihood scan curves for (left) the  mass fit and (right)  the angular fit. Blue curves show results with statistical uncertainties only, while red curves include the effect of systematic uncertainties, incorporated via convolution with a Gaussian distribution.
     }
    \label{fig:likelihoodscan_Massfit_angularfit}
 \end{figure}

The obtained yields of  \BuPLbar and \BuKSPi  decays in different  data categories are  given in  
Table~\ref{tab:branchratiocalculation:summary}. The profile-likelihood scans of the baseline mass fit and angular fit,  including the effects of systematic uncertainties,  are shown in Fig.~\ref{fig:likelihoodscan_Massfit_angularfit}. 

\begin{table}[htb]
\renewcommand{\arraystretch}{1.2} 
\caption{Branching fraction ratio $\mathcal{R}$ calculated in four data categories. The efficiency ratio is determined from simulation, and the yields are obtained from the baseline fit results. The first uncertainty on 
$\mathcal{R}$
 is statistical, and the second is systematic.
}
\small
%\footnotesize
\centering 
\begin{tabular}{lccccc}
\hline
Parameter        &TOS LL  &TOS DD  &TIS LL  &TIS DD    \\
\hline
$N(\KSPi)$    &12\,200 $\pm$ 140   &9250 $\pm$ 130   &5570 $\pm$ 100   &1480 $\pm$ 50    \\ 
$N(\PLbar)$   &\;\;\;40 $\pm$ 8   &26 $\pm$ 6   &17 $\pm$ 5   &\;\;\;\;5 $\pm$ 4     \\
$\frac{\epsilon(\KSPi)}{\epsilon(\PLbar)}$   &\;\;\;\;\;2.62 $\pm$ 0.14  &\;\;3.62 $\pm$ 0.18  &\;\;3.57 $\pm$ 0.21  &\;\;\;\;2.66 $\pm$ 0.24   \\
\hline
\multirow{2}{*}{$\mathcal{R} (10^{-2})$}  & $0.93 \pm 0.19 \pm 0.05$   & $1.11 \pm 0.28 \pm 0.07$   & $1.18 \pm 0.38 \pm 0.08$   & $0.95 \pm 0.82 \pm 0.14$     \\
& \multicolumn{4}{c}{$1.04 \pm 0.14 \pm 0.04$}  \\
%\bottomrule
%$\mathcal{R} (10^{-2})$ & $0.93 \pm 0.19 \pm 0.05$   & $1.11 \pm 0.28 \pm 0.07$   & $1.18 \pm 0.38 \pm 0.08$   & $0.95 \pm 0.82 \pm 0.14$     \\
%\hline
%$\mathcal{R} (10^{-2})$   &\multicolumn{4}{c}{$1.04 \pm 0.14 \pm 0.04$}  \\
\hline
\end{tabular}
\label{tab:branchratiocalculation:summary} 
\end{table}

 \clearpage

%\input{appendix}

% This should be taken out in the final paper
%\input{supplementary-app}

\newpage
\addcontentsline{toc}{section}{References}
\setboolean{inbibliography}{true}
\bibliographystyle{LHCb}
\bibliography{main,standard,LHCb-PAPER,LHCb-CONF,LHCb-DP,LHCb-TDR}

\newpage
% LHCb collaboration author list
% Data extracted on November 25th, 2025 at 2:28am for paper reference LHCb-PAPER-2025-044
\centerline
{\large\bf LHCb collaboration}
\begin
{flushleft}
\small
R.~Aaij$^{38}$\lhcborcid{0000-0003-0533-1952},
A.S.W.~Abdelmotteleb$^{58}$\lhcborcid{0000-0001-7905-0542},
C.~Abellan~Beteta$^{52}$\lhcborcid{0009-0009-0869-6798},
F.~Abudin{\'e}n$^{58}$\lhcborcid{0000-0002-6737-3528},
T.~Ackernley$^{62}$\lhcborcid{0000-0002-5951-3498},
A. A. ~Adefisoye$^{70}$\lhcborcid{0000-0003-2448-1550},
B.~Adeva$^{48}$\lhcborcid{0000-0001-9756-3712},
M.~Adinolfi$^{56}$\lhcborcid{0000-0002-1326-1264},
P.~Adlarson$^{86}$\lhcborcid{0000-0001-6280-3851},
C.~Agapopoulou$^{14}$\lhcborcid{0000-0002-2368-0147}
% ...

{\footnotesize \it

% ...
$^{14}$Universit{\'e} Paris-Saclay, CNRS/IN2P3, IJCLab, Orsay, France\\
% ...
$^{38}$Nikhef National Institute for Subatomic Physics, Amsterdam, Netherlands\\
% ...
$^{48}$Instituto Galego de F{\'\i}sica de Altas Enerx{\'\i}as (IGFAE), Universidade de Santiago de Compostela, Santiago de Compostela, Spain\\
% ...
$^{52}$Physik-Institut, Universit{\"a}t Z{\"u}rich, Z{\"u}rich, Switzerland\\
% ...
$^{56}$H.H. Wills Physics Laboratory, University of Bristol, Bristol, United Kingdom\\
% ...
$^{58}$Department of Physics, University of Warwick, Coventry, United Kingdom\\
% ...
$^{61}$School of Physics and Astronomy, University of Glasgow, Glasgow, United Kingdom\\
$^{62}$Oliver Lodge Laboratory, University of Liverpool, Liverpool, United Kingdom\\
% ...
$^{70}$Syracuse University, Syracuse, NY, United States\\
% ...
$^{86}$Department of Physics and Astronomy, Uppsala University, Uppsala, Sweden, associated to $^{61}$\\
% ...
\medskip
}
\end{flushleft}

\end{document}